\documentclass[twocolumn,showpacs,preprintnumbers,amsmath,amssymb]{revtex4}

\usepackage{graphicx}
\usepackage{dcolumn}
\usepackage{bm}

\begin{document}

\title{\large Interaction-Induced Spin Polarization in Quantum Dots}

\author{M. C. Rogge$^1$}
\email{rogge@nano.uni-hannover.de}
\author{E. R\"as\"anen$^2$}
\author{R.~J. Haug$^1$}
\affiliation{$^1$Institut f\"ur Festk\"orperphysik, Leibniz
Universit\"at Hannover, Appelstr. 2, 30167 Hannover,
Germany\\
$^2$Nanoscience Center, Department of Physics, University of
Jyv\"askyl\"a, FI-40014 Jyv\"askyl\"a, Finland}

\date{\today}

\begin{abstract}
The electronic states of lateral many electron quantum dots in
high magnetic fields are analyzed in terms of energy and spin. In
a regime with two Landau levels in the dot, several Coulomb
blockade peaks are measured. A zig-zag pattern is found as it is
known from the Fock-Darwin spectrum. However, only data from
Landau level 0 show the typical spin-induced bimodality, whereas
features from Landau level 1 cannot be explained with the Fock-Darwin
picture. Instead, by including the interaction effects within
spin-density-functional theory a good agreement between experiment
and theory is obtained. The absence of bimodality on Landau level
1 is found to be due to strong spin polarization.
\end{abstract}

\maketitle

Spin properties of semiconductor quantum dots (QDs) are of high
interest, as the spin of electrons captured in a QD could
be used to realize a quantum mechanical bit, the core of future
quantum information technologies \cite{Loss-98}. In few electron
QDs, the electronic spin has been successfully
implemented, manipulated and read (for a review see
\cite{Hanson-07}). However, in many electron systems, the dynamics
are still not well understood, especially in high magnetic fields.

\begin{figure}
 \includegraphics{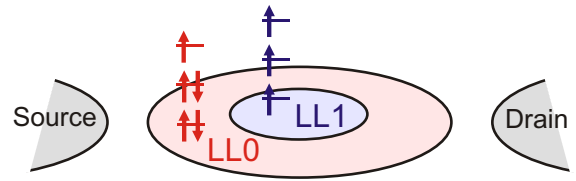}
 \caption{(color online) Schematic of a single lateral quantum dot connected to two leads (Source and Drain). In a perpendicular magnetic field, Landau levels (LL) appear with a specific spin polarization. Landau level 0 is located at the edge of the dot, Landau level 1 in the center. Arrows show the spin configuration within each Landau level.}
 \label{fig1}
\end{figure}

For QDs with many electrons, the simplest theoretical
approximation is done with the so called constant interaction (CI)
model \cite{Beenakker-91,Meirav-96}. This model uses
the single-electron states, most commonly those of a two-dimensional
harmonic potential (Fock-Darwin spectrum
\cite{Fock-28,Darwin-30}), and the many-body effects are included just by
a constant Coulomb repulsion energy. Despite its simplicity, many
features can be qualitatively explained using this model. Among
those are the formation of Landau levels (that allows to introduce
the QD filling factor $\nu$), the crossing of states leading to
zig-zag patterns, alternating spins and spin flips. Beyond the
CI model, self-consistent calculations successfully described the
regularity of zig-zag patterns and the electron densities within
certain Landau levels \cite{McEuen-92}. It was found that Landau
levels form conductive rings in the dot. Especially for $4>\nu>2$,
a central region is formed with Landau level 1 (LL1) and an outer
ring with Landau level 0 (LL0), as schematically shown in Fig.
\ref{fig1}.

Further investigations on spin blockade and Kondo
effect helped to gain knowledge about the spin configuration
in the lowest Landau levels (see, e.g.,
Refs.~\cite{Stopa-03,Keller-01,Ciorga-00,Rogge-06,Rogge-06b,spindroplet}).
Calculations by Wensauer et al. \cite{Wensauer-03} predicted a
spin polarization in LL0 at the $\nu=2$ border as a function of
the electron number. This was also found experimentally
\cite{Ciorga-02,Rogge-06}. In addition, collective spin polarization
on the highest occupied Landau level has been found in
many-electron calculations~\cite{nu52,spindroplet,spindroplet2}
indirectly supported by experimental data~\cite{Ciorga-00}.
However, despite all these results, the level spectrum of
many-electron QDs still leaves open questions, and further
measurements are required.

Here we concentrate on the spin configuration of both lowest
Landau levels in the $4>\nu>2$ regime. We present conductance
measurements on a lateral QD with approximately 50 electrons in a
perpendicular magnetic field. A zig-zag pattern of electronic
states is found and it is compared with many-body calculations
within spin-density-functional theory~\cite{dft} (SDFT). While the
states in LL0 show a striking bimodality that can roughly be
explained with alternating spin states using the CI model, the
states of LL1 do not at all follow this model. However, the
behavior can be understood with the SDFT calculations showing
different spin configurations in the two involved Landau levels as
illustrated in Fig. \ref{fig1}. While LL0 shows a regular filling
of orbitals with alternating spin-up and spin-down electrons as
expected, LL1 is spin-polarized with spin-up electrons only.

The sample is made on a GaAs/AlGaAs heterostructure using local
anodic oxidation \cite{Ishii-95,Held-98,Keyser-00} and electron
beam lithography. To allow electronic transport, the QD is connected
to two leads, source and drain, as schematically shown in Fig. \ref{fig1}.
A side gate is used to tune the QD
potential. Details about the sample preparation and device
properties can be found in Ref. \cite{Rogge-04}.

\begin{figure}
 \includegraphics{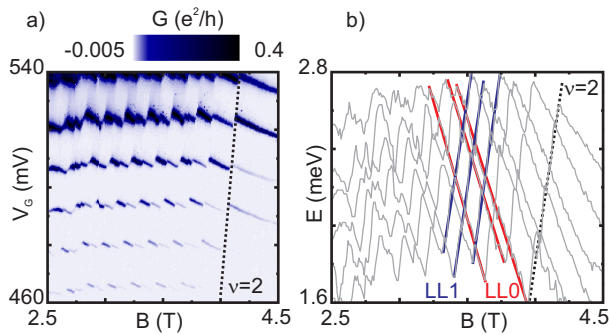}
 \caption{(color online) (a) Differential conductance $G$ as a function of magnetic field $B$ and gate voltage $V_G$. Six Coulomb peaks are visible showing a pronounced zig-zag pattern. (b) The peak positions are transferred to an energy scale with the Coulomb energy removed. A cross pattern is found with states going down in energy with increasing field (LL0) and states going up in energy (LL1).}
 \label{fig2}
\end{figure}

Figure \ref{fig2} (a) shows a measurement of the magnetotransport
of the system. The differential conductance $G$ is plotted as a
function of the perpendicular magnetic field $B$ and gate voltage
$V_G$. Six Coulomb peaks are visible. The positions of these peaks
reflect the chemical potentials of ground-state transitions. These
positions can be roughly understood using the CI model. According
to this model, the potentials include a fixed Coulomb repulsion
energy due to the electrostatic interaction of the electronic
charge and a term due to the single particle excitation spectrum.
The single-particle problem is usually calculated describing the
confinement of the QD as a two-dimensional harmonic potential,
which leads to the Fock-Darwin spectrum\cite{Fock-28,Darwin-30}.
In addition, the electronic spin is included via the Zeeman term
$g^*\mu_BBs_z$ with the gyromagnetic ratio $g^*$, the
Bohr-magneton $\mu_B$ and the spin quantum number $s_z=\pm 1/2$.
According to this model, the pronounced zig-zag structures appear
as with increasing magnetic field different states of the
excitation spectrum are energetically favored.

The bare excitation spectrum is made visible by removing the
Coulomb energy from the chemical potentials. This is done in Fig.
\ref{fig2} (b). The peak positions of the original Coulomb peaks
are transferred to an energy scale with the Coulomb energy removed
(we did actually remove a gate voltage dependent Coulomb energy,
as the electrostatic electron-electron interaction is reduced with
increasing $V_G$ due to changes in the size of the dot). Now the
states from the excitation spectrum can be followed over several
Coulomb peaks (several electron numbers). As a result, a pattern
is found with states going up in energy (some marked blue) and
states going down in energy (some marked red). These states can be
interpreted in terms of Landau levels. States with negative slopes
are due to transport via LL0, states with positive slopes are due
to transport via LL1. According to the Fock-Darwin model with
spin, for each Landau level there should be a pairing of every two
lines. Two adjacent peaks use the same orbital state with opposite
spin. Indeed such a pairing is visible for LL0 but not for LL1.

\begin{figure}
 \includegraphics{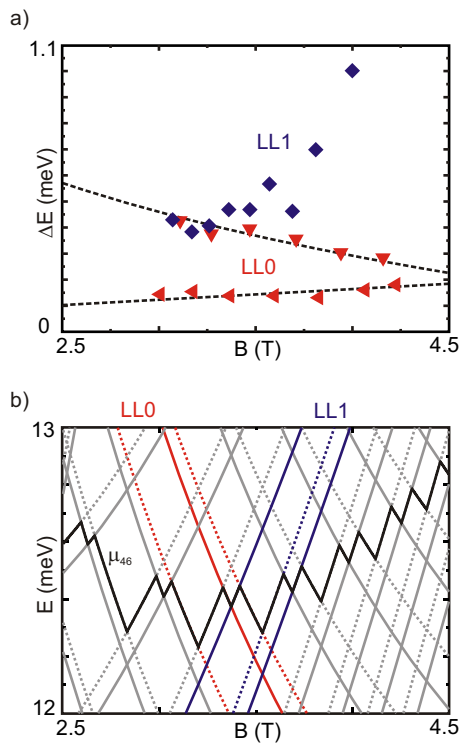}
 \caption{(color online) (a) Measured energetic peak distances for states in
   LL0 (triangles) and in LL1 (diamonds). The distances for LL0 show a
   bimodal behavior due to spin and can be fitted with the Fock-Darwin
   model (dashed lines, $\hbar\omega_0=1.83$~meV, $g^*=-0.71$). The
   Fock-Darwin spectrum using these parameters is shown in (b). Values
   for LL1 [diamonds in (a)] cannot be explained with the Fock-Darwin spectrum.}
 \label{fig3}
\end{figure}

In order to analyze this behavior in detail, for each Landau Level
the energy distances between adjacent peaks are plotted in Fig.
\ref{fig3} (a). The distances for LL0 (triangles) show the
expected bimodal behavior due to the electronic spin. The data can
be fitted (dashed lines) using the Fock-Darwin model with
confinement strength $\hbar\omega_0=1.83$~meV and $g^*=-0.71$. The
corresponding Fock-Darwin spectrum using these parameters is shown
in Fig. \ref{fig3} (b) around the chemical potential $\mu_{46}$
(black trace). Spin-up and spin-down states are shown as solid and
dotted lines, respectively. However, the large negative value for
the $g^*$-factor (the sign is given by the peak amplitudes due to
spin blockade \cite{Ciorga-00}) is a first hint, that the
description with the CI model is incomplete. Due to the
confinement potential of the dot, the $g^*$-factor should be more
positive than the one of $-0.44$ for bulk GaAs. And finally, the
description with the CI model is completely insufficient, when
data from LL1 are included. Within the Fock-Darwin model, the
energetic distances do not depend on Landau level, but they should
be identical for LL0 and LL1. This is obviously not the case. In
contrast to the Fock-Darwin spectrum [Fig. \ref{fig3} (b)] the
measured values for LL1 [diamonds in Fig. \ref{fig3} (a)] do not
show a bimodal behavior. Moreover, the values are larger as for
LL0 and increase nonlinearly with magnetic field. Thus a more
sophisticated approach is needed including electron-electron
interactions that go beyond the CI model.

In the many-body approach we consider the $N$-electron Hamiltonian
\begin{eqnarray}
H & = & \frac{1}{2m^*}\sum^N_{i=1}\left[\mathbf{p}_i+e\mathbf{A}
({\mathbf r}_i)\right]^2
+\sum^N_{i<j}\frac{e^2}{4\pi\epsilon_0\epsilon|{\mathbf r}_i-{\mathbf
    r}_j|} \nonumber \\
& + & \sum^N_{i=1}\left[V_{\rm ext}({\mathbf r}_i)+E_{z,i}\right],
\label{hamiltonian}
\end{eqnarray}
where ${\mathbf A}$ is the external vector potential (in symmetric
gauge) of the homogeneous, perpendicular magnetic field ${\mathbf
  B}=B{\hat z}$, and the last two terms correspond to the external
potential in the harmonic approximation $V_{\rm
  ext}(r)=m^*\omega_0^2 r^2/2$ with $\hbar\omega_0=4$ meV, and the Zeeman energy
$E_z=g^*\mu_BBs_z$,
respectively. We apply here the conventional effective-mass
approximation with the GaAs material parameters:
$m^*=0.067$ and $\epsilon^*=12.4$. For the gyromagnetic ratio we
have chosen $g^*=-0.30$.

We solve the ground-state energies associated with the $N$-electron
Hamiltonian~(\ref{hamiltonian}) by applying the SDFT~\cite{dft} in the
collinear-spin representation. We note that the external
vector potential is retained in the corresponding Kohn-Sham Hamiltonian.
To approximate the exchange-correlation
energy $E_{\rm xc}$ we use the local spin-density approximation
(LSDA) with a parametrization of the correlation energy in the homogeneous
2D electron gas by Attaccalite
{\em et al.}~\cite{attaccalite} For total-energy calculations on QDs, SDFT
with the LSDA has been shown to be a reliable scheme in comparison with quantum
Monte Carlo calculations, even in relatively high magnetic
fields~\cite{LDAtesting,spindroplet,spindroplet2}. Furthermore, it has been
found that the current-SDFT~\cite{vignale} (with the local-vorticity
approximation) does not lead to a considerable improvement
over the SDFT results~\cite{LDAtesting}. In the numerical calculations
we apply the {\tt octopus} code package~\cite{octopus}.

\begin{figure}
 \includegraphics{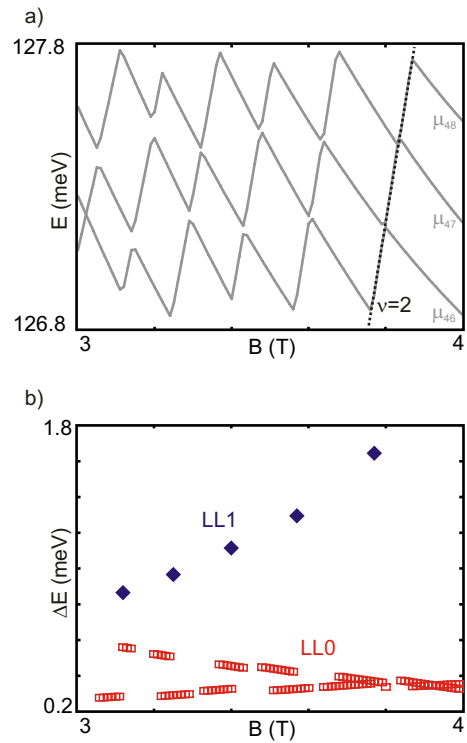}
 \caption{(color online) Results of the SDFT calculations:
(a) chemical potentials $\mu_{46}$, $\mu_{47}$ and $\mu_{48}$ without
Coulomb energy. As in the experiment, a cross pattern is found.
(b) Energetic distances for the two Landau levels. The data
qualitatively fit to the measurements. For LL0 a bimodal behavior
is found (squares). The values for LL1 (diamonds) increase with
increasing field and do not show a bimodality.}
 \label{fig4}
\end{figure}

Figure~\ref{fig4}(a)
shows the chemical potentials $\mu_N=E_N-E_{N-1}$ of the SDFT
calculations for $N=46\ldots 48$. The resulting energy-level
distances extracted from the chemical potentials are shown
in Fig.~\ref{fig4}(b). For LL1 we find a clear single mode,
whereas LL0 has a bimodal behavior as a function of $B$.
This is in a good qualitative agreement
with the experimental result in Fig.~\ref{fig3}(a).

\begin{figure}
 \includegraphics{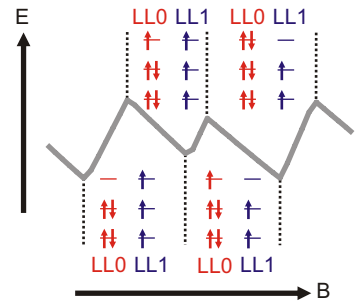}
 \caption{(color online) Spin configuration resulting from the
SDFT calculations for a typical section of the zig-zag pattern.
While electrons in LL0 are filled in regularly with alternating
spin, LL1 is spin polarized with electrons having spin up only.}
 \label{fig5}
\end{figure}

To analyze the physics behind the energy-level spacings, we
focus in the following on the spin configurations
given by the SDFT calculations. Their connection to the
chemical potentials is shown in Fig.~\ref{fig5}.
A typical portion of a Coulomb peak in the zig-zag regime is shown
together with the spin configurations for both Landau levels for
the adjacent regions of constant electron number. LL0 shows
alternating spin filling resulting in the bimodality of peak
sections with negative slopes. The bimodality is due to the energy
difference in adding an electron with spin up to a new orbital
compared with adding an electron with spin down to a half-filled
orbital. We note that the energy difference decreases as a
function of $B$. This is due to both the Zeeman effect and the
fact that the increasing magnetic confinement brings the energy
states in LL0 closer to each other, so that the energy in
occupying a new orbital decreases with respect to the filling of a
half-filled orbital (costing Coulomb energy).

In contrast with the situation for LL0, the higher Landau level
(LL1), corresponding to peak sections with positive slopes
in Figs.~\ref{fig4} and \ref{fig5}, is
completely spin-polarized with spin up.
Thus, the filling mechanism is always the same and no
bimodality is found. The single mode goes up in energy
due to the effective increase in the level spacings as
a function of $B$, i.e., LL1 becomes more compact,
so that the filling of the spin-up orbitals
becomes energetically more costly.

We point out that the qualitative features in Figs.~\ref{fig4} and
\ref{fig5} are stable with respect to the values chosen for the
gyromagnetic ratio. The reason behind the strong spin polarization
of LL1 is the high density of states close to the Fermi energy
leading to collective, local ferromagnetism familiar from the
Stoner effect~\cite{stoner} and Hund's rule.
Energetically, this
many-body transition corresponds to gaining electronic exchange
at the expense of kinetic energy in occupying higher levels.
Similar behavior corresponding to the formation of ``spin-droplets''
has been predicted by quantum Monte Carlo and SDFT calculations
on QDs, indirectly supported by the experimental data of Coulomb-blockade
peak positions~\cite{spindroplet}. It
was concluded that the spin-droplet formation requires a relatively
large number of electrons $(N\gtrsim 30)$ and considerable strength
of electron-electron interactions. These conditions are fulfilled
by our device, and hence the Landau-level distances visible in our
excitation spectra can be considered as a direct and distinct evidence
of the spin polarization on the highest occupied Landau level.

To conclude, we analyzed the spin configuration of a single
quantum dot containing approximately 50 electrons. In high magnetic
fields electronic states were investigated for the two lowest
Landau levels. For LL0 distances in the excitation spectra
show a bimodal behavior, while for LL1 only one mode appears. This
is explained with two different spin configurations in the two
Landau levels. While there is no spin polarization in LL0, LL1 is
completely spin polarized. This is a direct confirmation of the
interaction-induced collective spin polarization in quantum dots.

This work has been supported by BMBF via nanoQUIT, by QUEST, and
by the Academy of Finland.






\end{document}